\documentclass[preprint,aps,prd,groupedaddress]
              {revtex4}%
\usepackage{graphicx}
\usepackage{amsmath}
\usepackage{bm}
\usepackage{color}
\usepackage{relsize}
\RequirePackage{xspace}
\begin{document}


\title{\mbox{Oscillating scalar fields in extended quintessence}\\[10pt]
}

\author{Dan Li,$^{1,2}$ Shi Pi,$^{3}$ and Robert J. Scherrer$^{2}$}\affiliation{{\footnotesize $^1$~School of Science, Zhejiang University of Science and Technology(ZUST), 318 LiuHe Road, HangZhou, China, 310023\\
$^2$Department of Physics and Astronomy, Vanderbilt University, Nashville, TN, U.S., 37235\\
$^3$Center for Gravitational Physics, Yukawa Institute for Theoretical Physics, Kyoto University, Kyoto 606-8502, Japan}}




\begin{abstract}
We study a rapidly-oscillating scalar field with potential $V(\phi) = k|\phi|^n$ nonminally coupled
to the Ricci scalar $R$ via a term of the form $(1- 8 \pi G_0 \xi \phi^2) R$ in the action.
In the weak coupling limit, we calculate the effect of the nonminimal coupling
on the time-averaged equation of state parameter $\gamma = (p + \rho)/\rho$.
The change in $\langle \gamma \rangle$
is always negative for $n \ge 2$ and
always positive for $n < 0.71$ (which includes the case where the oscillating
scalar field could serve as dark energy), while it can be either positive or negative for intermediate
values of $n$.
Constraints on the time-variation of $G$ force
this change to be infinitesimally small at the present time whenever the scalar field
dominates the expansion, but constraints in the early universe are not as stringent.
The rapid oscillation induced in $G$ also produces an additional contribution
to the Friedman equation that behaves like an effective energy density
with a stiff equation of state, but we show
that, under reasonable assumptions, this effective energy density is always smaller than the density of the scalar
field itself.
\end{abstract}
\preprint{YITP-17-103}
\maketitle
\section{Introduction}

Rapidly oscillating scalar fields have long been of interest in cosmology.  They were first systematically
explored by Turner \cite{Turner} and later reexamined by many others, often in the context of
dark energy \cite{liddle,sahni,hsu,masso,gu,dutta}.
These earlier studies took
the scalar field to be minimally coupled.
In this paper, we extend these results to an oscillating scalar field that is
nonminimally coupled to the curvature scalar $R$ via a term of the form $(1- 8 \pi G_0\xi \phi^2) R$ in the action.
Models with this particular coupling have been dubbed
``extended quintessence," and they have been extensively studied with
a variety of different potentials \cite{Uzan,Amendola,Perrotta,Holden,Baccigalupi,Chen,Chiba}.

In this paper, we examine the behavior of a rapidly-oscillating extended quintessence field for
which the oscillation frequency $\nu$ is much larger than the Hubble expansion rate $H$.
There have been several previous studies related to such models.  Luo and Su \cite{Luo} performed a numerical
study of oscillating extended
quintessence models for which $\nu \sim H$.  El-Nabulsi \cite{ElNabulsi}
performed a similar numerical study for the particular
case of conformal coupling.  Perivolariopoulos \cite{Perivolaropoulos} examined Brans-Dicke theories with an oscillating scalar.
The previous work most similar to our own is Ref. \cite{Steinhardt}, which examined nonminimally coupled models with the same
coupling as that examined here, and numerically calculated the equation of state for these models in the Einstein frame.
Here we work in the physical frame, and derive analytically the equation of state parameter in the limit of weak coupling.
We also derive updated observational constraints on such models.  While we have used the term ``extended quintessence"
in accordance with earlier nomenclature for this type of nonminimally-coupled scalar field, we do not intend to confine
our discussion to models in which the scalar field provides the dark energy; instead, we are interested in the general
cosmological behavior of these models.

In Sec. II, we review the standard results for minimally-coupled oscillating scalar fields.
In Sec. III, we derive the basic results for oscillating extended quintessence, particularly the change in the
equation of state parameter due to the nonminimal coupling. 
In Sec. IV, we present observational constraints on these models.
We discuss our conclusions briefly in Sec. V.

\section{Review of Minimally-coupled Oscillating Scalar Fields}
First, recall the behavior of a minimally-coupled scalar field $\phi$ oscillating in the potential
\begin{equation}
\label{Vphi}
V = k |\phi|^n.
\end{equation}
For a minimally-coupled scalar field, the pressure and density are given by
\begin{eqnarray}
\label{pphi}
p_\phi &=& \frac{\dot \phi^2}{2} - V(\phi),\\
\label{rhophi}
\rho_\phi &=& \frac{\dot \phi^2}{2} + V(\phi),
\end{eqnarray}
respectively, where the dot denotes the time derivative.  The quantity of greatest interest is the evolution of the
scalar field energy density with respect to the scale factor $a$, which is given by
\begin{equation}
\frac{d\ln\rho_\phi}{d\ln a} = -3\gamma, 
\end{equation}
where $\gamma$ is defined as
\begin{equation}
\label{gamma}
\gamma = \frac{p_\phi + \rho_\phi}{\rho_\phi}.
\end{equation}
The equation of motion for $\phi$ is
\begin{equation}
\label{motionq}
\ddot{\phi}+ 3H\dot{\phi} + \frac{dV}{d\phi} =0,
\end{equation}
where $H$ is the Hubble parameter, which depends on the total density $\rho_T$ as
\begin{equation}
\label{H}
H = \left(\frac{\dot{a}}{a}\right) = \sqrt{8 \pi G \rho_T/3}.
\end{equation}

Following Turner \cite{Turner}, 
we note that in the limit where the oscillation frequency is much larger than $H$, the density will evolve
slowly relative to the oscillation timescale.  In that case, Eqs. (\ref{pphi}), (\ref{rhophi}), and (\ref{gamma})
give
\begin{equation}
\label{gammamean}
\langle \gamma \rangle = \frac{\langle \dot \phi^2 \rangle}{\rho_\phi}.
\end{equation}
where the averages are taken over one oscillation period.  Let $\phi_\text{min}$ and $\phi_\text{max}$ be the minimum and
maximum values for $\phi$.  By symmetry $\phi_\text{max} = - \phi_\text{min} \equiv \phi_m$, and the value of $\rho_\phi$, which is
effectively constant over one oscillation period, is just $\rho_\phi  = V(\phi_m) \equiv V_m$.  Then the period-averaged
value of $\gamma$ given by Eq. (\ref{gammamean}) can be written as
\begin{eqnarray}
\langle \gamma \rangle &=& \frac{1}{V_m} \frac{\int {\dot \phi^2} dt}{\int dt},\\
&=& \frac{1}{V_m}\frac{\int \dot \phi ~d\phi}{\int (1/\dot \phi) d\phi},\\
&=& 2 \frac{\int_{-\phi_m}^{\phi_m} [1-V(\phi)/V_m]^{1/2} d\phi}{\int_{-\phi_m}^{\phi_m} [1-V(\phi)/V_m]^{-1/2} d\phi}.
\end{eqnarray}
For a power law of the form given by Eq. (\ref{Vphi}), the integrals can be evaluated exactly, yielding the main
result of Ref. \cite{Turner}:
\begin{equation}
\label{gamma0}
\langle \gamma \rangle = 2n/(n+2).
\end{equation}
This result implies that
$\rho_\phi$ scales as $\rho_\phi \propto a^{-6n/(n+2)}$, so the cases $n=2$ and $n=4$ correspond to behavior resembling nonrelativistic matter ($\rho \propto a^{-3}$) and
radiation ($\rho \propto a^{-4}$), respectively.  We can also derive the dependence
of $\phi_m$ on the scale factor, a result that will be used later.  From
$\rho_\phi \propto a^{-6n/(n+2)}$ and $\rho_\phi = k |\phi_m|^n$, we get
\begin{equation}
\label{phim}
\phi_m \propto a^{-6/(n+2)}.
\end{equation}
(See Ref. \cite{GP} for a more exact treatment).

It is actually rather difficult to produce realistic models in which the oscillating scalar field serves as dark energy.  A simple-minded application
of Eq. (\ref{gamma0}) with $\langle \gamma \rangle \approx 0$, as required by observations, would necessitate
a very flat potential ($n \ll 1$).  This can be avoided by using potentials of an unusual shape \cite{masso},
but Johnson and Kamionkowski have argued that any rapidly-oscillating scalar field with
$\langle \gamma \rangle < 1$ will be unstable to the growth of inhomogeneities \cite {JK}.

\section{Oscillating Extended Quintessence}

We now extend the calculation of the previous section to
nonminimally coupled quintessence. The action for the scalar field with potential $V(\phi)$ in the Jordan frame is
\begin{equation}
\label{action}
S_{\phi}=-\frac{1}{2}\int \left[Z(\phi)\partial_\mu\phi\partial^\mu\phi-\frac{F(\phi)
R}{8 \pi G_0}+2V(\phi)\right]\sqrt{- g}~d^4x,
\end{equation}
where $G_0$ is the bare gravitational constant.
We examine the model discussed previously in Refs. \cite{Perrotta,Baccigalupi,Chen}, namely
we choose $Z(\phi)=1$ and $F(\phi)=1- 8 \pi G_0 \xi \phi^2$.
Note that we follow the convention of Ref. \cite{Amendola} in including a factor of $8 \pi G_0$
in the definition of $F(\phi)$, as opposed to, e.g., the convention of Refs. \cite{Perrotta,Baccigalupi,Chen}.
For the potential $V(\phi)$, we take, as in the minimally-coupled case,
\begin{equation}
\label{Vphi2}
V(\phi) = k|\phi|^n.
\end{equation}
We assume that the field undergoes rapid oscillation about $\phi = 0$, with oscillation
frequency $\nu \gg H$.  This oscillation is superimposed on a slow decay in the oscillation amplitude.
Hence, the case $\phi = 0$, $G  = G_0$ corresponds not to the present day but the asymptotic future.
The present-day value of $G$ is given by $G = G_0 \langle [1 - 8 \pi G_0 \xi \phi^2)]^{-1} \rangle$,
where the average is taken over an oscillation period.

We adopt the flat Friedmann-Robertson-Walker universe with the line element $ds^2=-dt^2+a^2(t)[\delta_{ij}dx^idx^j]$.
The equation of motion for $\phi$ is then
\begin{equation}
\label{EoM}
\ddot\phi+3H\dot \phi +\frac{\partial V}{\partial \phi}+ \xi R \phi = 0.
\end{equation}
The Ricci scalar $R$ can be expressed in terms of Hubble parameter as:
\begin{equation}
R=6(2H^2+\dot H).
\end{equation}
It is then straightforward to derive the scalar field density and pressure \cite{Uzan}:
\begin{eqnarray}
\label{density}
\rho_{\phi}&=&\frac{1}{2}\dot\phi^{2}+V(\phi)+6\xi H
\phi\dot\phi+ 3\xi H^2\phi^2\\
\label{pressure}
p_{\phi}&=&\frac{1}{2}\dot\phi^{2}-V(\phi)-\xi\left((2{\dot H}+3{H}^2)\phi^2
+4{H}\phi\dot\phi+2\phi\ddot\phi+2\dot\phi^{2}\right).
\end{eqnarray}
In what follows we focus only on the weak coupling regime:
$0<\xi \ll1$, in order to make our calculations tractable, and all quantities of interest will be expanded
to linear order in $\xi$.

In the limit of interest, $\nu \gg H$,
the first and third terms in Eq. (\ref{EoM}) are dominant, and (to lowest order in $\xi$) the field undergoes
symmetric oscillations around $\phi = 0$.
Then, from Eqs. (\ref{EoM}), (\ref{density}), and (\ref{pressure}) the adiabatic index of the scalar field averaged over one
oscillation period is,
to linear order in $\xi$,
\begin{equation}
\label{gammaNMC}
\langle \gamma \rangle = \frac{\langle \rho_\phi+p_\phi \rangle }{\rho_\phi}=\frac{1}{\rho_\phi}
\biggl[\langle \dot \phi ^2 \rangle +2\xi(\langle nV(\phi) \rangle - \langle \dot\phi^2 \rangle )+8\xi H
\langle \phi\dot\phi\rangle - 2 \xi \dot H \langle \phi^2 \rangle \biggr].
\end{equation}
We first note that $\langle \phi\dot\phi \rangle=0$.  Furthermore,
the period-averaged kinetic term is equal to the period-averaged potential term up to order $\xi$: $\langle nV(\phi) \rangle -
\langle \dot\phi^2 \rangle \sim
O(\xi)$ \cite{gu}.

The value of $\rho_\phi$ in the denominator of Eq. (\ref{gammaNMC}) is modified from its value in the minimally-coupled
case.  For the field oscillating between $\phi = \phi_\text{min}$ and $\phi = \phi_\text{max}$, we define, in a similar manner to
the discussion in previous
section, $\phi_m \equiv \phi_\text{max} =  - \phi_\text{min}$, and $V_m = V(\phi_m)$.
Then evaluating Eq. (\ref{density}) at either extremum,
we obtain, for the oscillating scalar field,
\begin{equation}
\rho_\phi = V_m + 3 \xi H^2 \phi_m^2.
\end{equation}
Then our expression for the adiabatic index becomes
\begin{equation}
\label{gammaNMC2}
\langle \gamma \rangle = \frac{\langle \dot \phi ^2 \rangle  - 2\xi \dot H \langle \phi^2 \rangle}{V_m + 3 \xi H^2 \phi_m^2},
\end{equation}
which, to linear order in $\xi$, is
\begin{equation}
\label{gammaNMC3}
\langle \gamma \rangle = \frac{1}{V_m} \left[1 - \frac{3 \xi H^2 \phi_m^2}{V_m}\right] \langle \dot \phi^2 \rangle
- \frac {2 \xi \dot H}{V_m} \langle \phi^2 \rangle.
\end{equation}
We now proceed to calculate $\langle \dot \phi^2 \rangle$ and $\langle \phi^2 \rangle$ using methods similar to those in the previous
section.

First consider $\langle \dot \phi^2 \rangle$.  As in the previous section, we have
\begin{equation}
\label{phidot2}
\langle \dot \phi^2 \rangle = \frac{\int \dot \phi ~d\phi}{\int (1/\dot \phi) d\phi},
\end{equation}
but we now define an effective potential given by
\begin{equation}
\label{Veff}
V_\text{eff}(\phi) = V(\phi) + 3 \xi H^2 \phi^2.
\end{equation}
Treating Eq. (\ref{density}) as a quadratic equation in $\dot \phi$, we obtain
\begin{eqnarray}
\dot \phi &=& - 6 \xi H \phi \pm \sqrt{2(\rho_\phi - V_\text{eff}) + 36 \xi^2 H^2 \phi^2},\\
\label{dotphi}
&=& - 6 \xi H \phi \pm \sqrt{2(\rho_\phi - V_\text{eff})}~~~{\rm(to ~linear ~order ~in ~\xi)}.
\end{eqnarray}
Then to linear order in $\xi$, we also have
\begin{equation}
\label{dotphi-1}
1/\dot \phi = \frac{1}{\sqrt{2(\rho_\phi - V_\text{eff})}} + \frac{3 \xi H \phi}{(\rho_\text{eff}-V_\text{eff})}.
\end{equation}

When we substitute the expressions from Eqs. (\ref{dotphi}) - (\ref{dotphi-1}) into Eq. (\ref{phidot2}), the terms
linear in $\phi$ vanish in both integrals, leaving
\begin{equation}
\label{phidot2final}
\frac {1}{V_m} \langle \dot \phi^2 \rangle = \frac{1}{V_m}\frac{\int [2(\rho_\phi - V_\text{eff})]^{1/2}d\phi}
{\int [2(\rho_\phi - V_\text{eff})]^{-1/2} d\phi}.
\end{equation}
However, this is just the value of $\gamma$ for the minimally-coupled case
with a potential given by Eq. (\ref{Veff}), i.e., a power-law plus a small
(different) power-law
correction.  The value of $\gamma$ for this particular case was calculated in Ref. \cite{Turner}; using that result, we derive:
\begin{equation}
\label{phidot2finalfinal}
\frac {1}{V_m} \langle \dot \phi^2 \rangle = \frac{2n}{n+2} + \frac{3 \xi H^2 \phi_m^{2-n}}{k} \frac{4(2-n)(3-n)}{(6-n)(n+2)}
\frac{\Gamma(\frac{1}{2}+\frac{1}{n}) \Gamma(\frac{3}{n}-1)}{\Gamma(\frac{1}{n}) \Gamma(\frac{3}{n} - \frac{1}{2})}. 
\end{equation}

Now consider the second term in Eq. (\ref{gammaNMC3}).  Since it is multiplied
by $\xi$, we can neglect any order-$\xi$ corrections to $\langle \phi^2
\rangle$.  To zeroth order in $\xi$, we then have simply
\begin{equation}
\label{phi2int}
\langle \phi^2 \rangle =
\frac{\int \phi^2 [2(\rho_\phi - V)]^{-1/2}d\phi}{\int[2(\rho_\phi -
V)]^{-1/2}d\phi},
\end{equation}
where the $V(\phi)$ appearing in this equation is the zeroth order potential
given by Eq. (\ref{Vphi2}), rather than $V_\text{eff}$.  The integrals in Eq. (\ref{phi2int}) yield
\begin{equation}
\label{phi2final}
\langle \phi^2 \rangle = \phi_m^2 \frac{\Gamma(\frac{3}{n})\Gamma(\frac{1}{n} + \frac{1}{2})}
{\Gamma(\frac{1}{n})\Gamma(\frac{3}{n} + \frac{1}{2})}
\end{equation}
We combine Eqs. (\ref{gammaNMC3}), (\ref{phidot2finalfinal}), and (\ref{phi2final}) to obtain a final expression
for $\langle \gamma \rangle$.  Because the scalar field oscillations correspond to a timescale much shorter than $H^{-1}$, we can take both
$H$ and $\dot H$ to be constant, and given by $H^2 = 8 \pi G_0 \rho_T/3$ and $\dot H = -4 \pi G_0 (\rho_T + p_T)$, where
$\rho_T$ and $p_T$ are the total (scalar field plus background radiation or matter) density and pressure, respectively.

Finally, since $V_m = \rho_\phi - 3 \xi H^2 \phi_m^2$, and dropping terms linear in $\xi$ when plugging back into Eq. (\ref{gammaNMC3}),
we obtain
\begin{equation}
\label{gammafinal}
\langle \gamma \rangle = \frac{2n}{n+2}
+ 8 \pi G_0 \xi \phi_m^2 \frac{\rho_T}{\rho_\phi}
\left[
\left(\frac{2(2-n)}{n+2}+\gamma_T\right) K(n) -
\left(\frac{2n}{n+2}\right)\right],
\end{equation}
where $\gamma_T \equiv (p_T+\rho_T)/\rho_T$, and $K(n)$ is defined as
\begin{equation}
K(n) \equiv \frac{\Gamma(\frac{3}{n})\Gamma(\frac{1}{n} + \frac{1}{2})}
{\Gamma(\frac{1}{n})\Gamma(\frac{3}{n} + \frac{1}{2})}.
\end{equation}
Note that $K(n)$ is a slowly-varying function of $n$:  for $0 < n < 10$,
we have $1/\sqrt{3} > K(n) > 0.4$.
Eq. (\ref{gammafinal}) is our main result.
It gives, to lowest order in $\xi$, the change in the equation of state
parameter for an oscillating scalar field due to its nonminimal coupling.

Now consider the sign of the change in $\langle \gamma \rangle$ induced by the nonminimal coupling.  If
we define $\gamma_c$ to be given by
\begin{equation}\label{gammac}
\gamma_c = \frac{2n}{(n+2)K(n)} - \frac{2(2-n)}{n+2},
\end{equation}
then Eq. (\ref{gammafinal}) indicates that the change in $\langle \gamma \rangle$ is negative when
$\gamma_T < \gamma_c$ and positive when $\gamma_T > \gamma_c$.  
In general, the constraint $\gamma_T<2$ is reasonable for ordinary cosmic components. This can be justified, for instance, by assuming
that all the other components beyond $\phi$ obey the dominant energy condition (DEC), which gives $\gamma_i\leq2$ for each component
$i$. The physical significance of the DEC is that for a perfect fluid $i$,~$\gamma_i\leq2$ is equivalent to the causality condition
$c_{s(i)}^2=dp_i/d\rho_i\leq1$. These inequalities, together with the fact that $\gamma_\phi<2$ for all finite values of $n$, yield the constraint for the total equation of state parameter: $\gamma_T<2$. In this paper we will focus, for simplicity,
on this condition for $\gamma_T$, leaving aside the possibility that it may be violated
in ekpyrotic/cyclic models or other scenarios \cite{Khoury:2001wf,Steinhardt:2004gk, MV}.
Therefore, since $\gamma_c\geq2$ when $n \geq 2$, we see that the DEC bound $\gamma_T<2$ automatically guarantees $\gamma_T<\gamma_c$,
thus implying
that the change in $\langle \gamma \rangle$ is always negative when $n \geq 2$.  

In the opposite limit, we note that the weak energy condition (WEC) implies that $\gamma_T \ge 0$. Thus,
the change in $\langle \gamma \rangle$ will always be positive when $\gamma_c < 0$.  Using Eq. (\ref{gammac}), 
we find that the
latter condition is equivalent to $n < 0.71$. This corresponds (for the minimally-coupled case) to
$\langle \gamma \rangle < 0.5$.  In the intermediate regime, $0.71 < n < 2$, the change in $\langle \gamma
\rangle$ can be either positive or negative, depending on the value of $\gamma_T$.

Now consider a few illustrative examples.  For $n=2$, the behavior
of the minimally-coupled scalar field is dust-like ($\gamma = 1$), and Eq. (\ref{gammafinal}) gives
\begin{equation}
\label{gamma=1}
\langle \gamma \rangle = 1 + 8 \pi G_0 \xi \phi_m^2 \frac{\rho_T}{\rho_\phi}\left(\frac{1}{2}\gamma_T - 1\right).
\end{equation}
The other case of greatest interest is $n=4$, as it gives
rise to radiation-like ($\gamma = 4/3$)
behavior.  For this case, we obtain
\begin{equation}
\label{gamma=4/3}
\langle \gamma \rangle = \frac{4}{3} + 8 \pi G_0 \xi \phi_m^2 \frac{\rho_T}{\rho_\phi}(0.46\gamma_T - 1.64).
\end{equation}
As expected, both Eqs. (\ref{gamma=1}) and (\ref{gamma=4/3}) indicate that the 
effect of the nonminimal coupling for $\gamma_T < 2$  is to decrease the
value of the equation of state parameter relative to its minimally-coupled
value. 
These results are consistent with the numerical results in Ref. \cite{Steinhardt}, which show a similar decrease in
$\langle \gamma \rangle$ relative
to the minimally-coupled case.

Finally, consider the case where the oscillating scalar field serves as dark energy.  As noted in Sec. II, this is
difficult to achieve, but we include it here for completeness.  Current observations constrain the
dark energy equation of state parameter to be $\gamma_{DE} \ll 1$, which corresponds to $n \ll 1$ for an oscillating
scalar field.  Thus, the argument above indicates that the nonminimal coupling will increase
the value of $\langle \gamma \rangle$ relative to its minimally coupled value.

However, we will see in the next section that any changes in $\langle \gamma \rangle$ due
to the nonminimal coupling of the scalar field are sharply constrained by observations.

\section{Observational constraints}

The model examined here is constrained by observational limits on the time variation of $G$, which is
given by
\begin{equation}
\label{G}
G = G_0 / (1- 8 \pi G_0 \xi \phi^2).
\end{equation}
The time variation of $G$ tracks the time variation in $\phi$, which can be broken into two parts:  the rapid oscillation of $\phi$ with frequency $\nu$ much greater
than $H$, and the slow decay in the amplitude $\phi_m$.

Consider first the slow secular variation in $G$.  Measurements of the orbit of Mars give \cite{Will}
\begin{equation}
\label{Glimit}
\dot G/G < 1.7 \times 10^{-13} {\rm yr} ^{-1},
\end{equation}
at the present time.
To compare with our model, we need to average
Eq. (\ref{G}) over a single oscillation period.  Taking $8 \pi G_0 \xi \phi_m^2 \ll 1$, we can expand
Eq. (\ref{G}) to give
\begin{equation}
\langle G \rangle = G_0[1 + 8 \pi G_0 \xi \langle \phi^2 \rangle].
\end{equation}
Taking $\langle \phi^2 \rangle$ from Eq. (\ref{phi2final}), we obtain
\begin{equation}
\label{Gphim}
\langle G \rangle = G_0[1 + 8 \pi G_0\xi 
\phi_m^2 K(n)].
\end{equation}
Then to lowest order in $\xi$,
\begin{equation}
\langle \dot G\rangle/\langle G \rangle = 16 \pi G_0 \xi \phi_m \dot \phi_m K(n).
\end{equation}
Using $\rho_\phi = k \phi_m^n + O(\xi)$, we have $\dot \rho/\rho = n \dot \phi_m/\phi_m = -3 H \langle \gamma \rangle$,
with $\gamma$ given by Eq. (\ref{gamma0}). Then we obtain
\begin{equation}
\langle \dot G\rangle/\langle G \rangle = - \frac{96}{n+2} H \pi G_0 \xi \phi_m^2 K(n).
\end{equation}
For $H_0 = 70$ km sec$^{-1}$ Mpc$^{-1}$, the limit in Eq. (\ref{Glimit})
gives
\begin{equation}
\label{bound1}
8 \pi G_0 \xi \phi_m^2 < 0.0024 \frac{n+2}{12} \frac{1}{K(n)}\sim 10^{-3}.
\end{equation}
Since equation (\ref{Gphim}) gives us the present-day value of $G$, while
$G_0$ is the asymptotic value as the oscillating field decays to zero
amplitude, this limit tells us that $G$ will change by no more than
0.1\% in the asymptotic future.

This bound also limits the deviation of $\langle \gamma \rangle$ from its value in the minimally-coupled
case.  For example, consider the case $n=2$ in Eq. (\ref{gamma=1}). Combining this equation with
the bound from equation (\ref{bound1}) shows that the extended quintessence modification to the 
value of $\gamma$ is infinitesimally small, unless $\rho_\phi/\rho_T \lesssim 10^{-3}$, i.e., the scalar
field energy density is a tiny fraction of the total energy density in the universe.  Note, however,
that this bound applies only at the present day, as it is based on current measurements of $\dot G/G$.

While $\dot G/G$ cannot be measured directly in the early universe, it is possible to use big bang nucleosynthesis (BBN)
to constrain the total change in $G$ between the epoch of BBN ($T \sim 1$ MeV) and the present.  Refs. \cite{Copi,Bambi}
give
\begin{equation}
\frac{|G_\text{BBN} - G_\text{now}|}{G_\text{now}} < 0.2 - 0.3.
\end{equation}
(See also Ref. \cite{Barrow} for the specific case of Brans-Dicke models).
We can use Eq. (\ref{Gphim}) to translate this into a limit on the difference between $\phi_m^2$ at BBN
and $\phi_m^2$ at the present:
\begin{equation}
8 \pi G_0 \xi |\phi_{m\text{(now)}}^2 - \phi_{m\text{(BBN)}}^2| K(n) < 0.2 - 0.3.
\end{equation}
But from Eq. (\ref{phim}), we can conclude that $\phi_{m~now} \ll \phi_{m~BBN}$, so that our limit becomes
\begin{equation}
8 \pi G_0 \xi \phi_{m~BBN}^2 < (0.2 - 0.3)\frac{1}{K(n)}.
\end{equation}
This BBN bound is not as stringent as the limit on $\xi \phi_m^2$
at the present.  Comparing with Eq. (\ref{gammafinal}), we see
that the change in $\langle \gamma \rangle$ compared to the
minimally coupled case could be nonnegligible at the epoch of BBN, and there
are essentially no limits at earlier times.

For extended quintessence with a slowly varying scalar field,
solar-system limits on the Jordan-Brans-Dicke parameter generally provide
stronger constraints than the limits on the time-variation of $G$ \cite{Perrotta}.  However, this is
not the case for the rapidly-oscillating scalar fields considered here.
As noted in Refs. \cite{Steinhardt,Accetta}, modifications to standard general relativity
are undetectable on length scales above $\nu^{-1}$.  Thus, one can always postulate a sufficiently large $\nu$ to evade
both solar-system and laboratory constraints on these models.

Finally, the models examined here lead to a high-frequency oscillation in $G$ induced by the rapidly-oscillating scalar field.  While it might seem that
this could not have any observable effects, it was noted by Accetta and Steinhardt \cite{Accetta,Accetta2}
that the oscillation in $G$
modifies the Friedman equation in such a way as to produce an additional effective component of the energy density.
Specifically, when $G$ varies, the Friedman equation becomes \cite{Accetta}
\begin{equation}
\label{HandG}
H = \frac{1}{2} \frac{\dot G}{G} + \left[\frac{8 \pi G\rho_T}{3} + \frac{1}{4} \left(\frac{\dot G}{G}\right)^2 \right]^{1/2}.
\end{equation}
If the oscillation frequency $\nu$ satisfies $\nu \gg H$, then the first $\dot G/G$ term averages to zero, but $(\dot G/G)^2 \ne 0$.
Thus, this term contributes an effective energy density, $\rho_\text{eff}$, to the expansion of the universe
that is in addition to the contribution $\rho_\phi$ from the scalar field itself.  From Eq. (\ref{HandG}), this additional
effective energy density is given by
\begin{equation}
\rho_\text{eff} = \frac{3}{32 \pi G}\left(\frac{\dot G}{G}\right)^2.
\end{equation}
Then for our particular model, Eq. (\ref{G}) gives, to lowest order in $\xi$,
\begin{equation}
(\dot G/G)^2 = 256 \pi^2 G_0^2 \xi^2 \phi^2 \dot \phi^2.
\end{equation}
We can derive $\rho_\text{eff}$ by averaging this expression over an oscillation period:
\begin{eqnarray}
\rho_\text{eff} &=& 24 \pi G_0 \xi^2 \langle \phi^2 \dot \phi^2 \rangle,\nonumber\\
&=& 24 \pi G_0 \xi^2 \frac{\int \phi^2 \dot \phi d\phi}{\int (1/\dot \phi) d\phi},\nonumber\\
\label{rhoeff}
&=& 24 \pi G_0 \xi^2 \phi_m^2  \rho_\phi
\frac{\Gamma(\frac{3}{n})\Gamma(\frac{1}{2} + \frac{1}{n})}{\Gamma(\frac{1}{n}) \Gamma(\frac{3}{2} + \frac{3}{n})}
\end{eqnarray}
Since $\phi_m$ decreases with the expansion of the universe, $\rho_\text{eff}$ necessarily decays more rapidly than $\rho_\phi$.
To determine the variation of $\rho_\text{eff}$ with the scale factor, we note that, to lowest order in $\xi$,
$\rho_\phi$ scales as $\rho_\phi \propto a^{-6n/(n+2)}$ (Eq. \ref{gamma0}), while
$\phi_m \propto a^{-6/(n+2)}$ (Eq. \ref{phim}).
Then we obtain
\begin{equation}
\rho_\text{eff} \propto a^{-6}.
\end{equation}
Thus, $\rho_\text{eff}$ behaves like a component of the energy density with a stiff equation of state, independent of the value of $n$.

We can rewrite Eq. (\ref{rhoeff}) to obtain an expression for $\rho_\text{eff}/\rho_{\phi}$:
\begin{equation}
\frac{\rho_\text{eff}}{\rho_\phi} = (8 \pi G_0 \xi \phi_m^2)(\xi) \times O(1).
\end{equation}
All of the factors on the right-hand side are less than 1, so $\rho_\text{eff}/\rho_\phi < 1$, and
we can neglect the contribution of $\rho_\text{eff}$ to the cosmological expansion in comparison to the scalar field energy
density itself.  Note that tighter constraints on oscillations in $G$ can be derived from the orbits
of the planets in the Solar System when $\nu \sim $ yr$^{-1}$ \cite{Iorio}.

\section{Conclusions}

We have derived the change in the time-averaged equation of state parameter $\langle \gamma \rangle$ for an oscillating scalar field with a
nonminimal coupling of the form $(1-8 \pi G_0 \xi \phi^2)R$.  
Our results indicate that the effect of the nonminimal coupling is to decrease
$\langle \gamma \rangle$ below its value for a minimally-coupled oscillating scalar field for potentials
of the form $V(\phi) = k |\phi|^n$ when $n \ge 2$.  Conversely, the change in $\langle \gamma \rangle$
is always positive for $n < 0.71$ (which includes all power-law oscillating models for which the scalar field could serve as dark energy). For 
intermediate values of $n$, this change can be either positive or negative, depending on the value of $n$ and the total
equation of state parameter $\gamma_T$.
However, as noted in the previous section, current limits
on the time variation of $G$ constrain the present-day change in $\langle \gamma \rangle$ to be
negligible whenever the scalar field provides a substantial contribution to the total energy density.

These constraints are relaxed in the early universe.  While BBN also constrains the time-variation of $G$, these limits
are considerable weaker than present-day bounds.  Thus, the nonminimal coupling could provide a nonnegligible change in the scalar field equation of
state in the early universe even when the scalar field contributes substantially to the total energy density.

Another cosmological effect arises from the rapidly-oscillating value of $G$ in the Friedman equation.  While this
oscillation yields an effective
energy density scaling as $a^{-6}$, we have shown that this effective energy density will always be dominated by the density
of the scalar field itself.

\begin{acknowledgments}
We thank Yong Hui Li and Yi-fu Cai for helpful discussions. SP is supported by MEXT KAKENHI No. 15H05888.
\end{acknowledgments}

\end{document}